\documentclass[10pt,a4paper]{article}
\usepackage{graphicx}

\usepackage{authblk}
\usepackage{fullpage,amsfonts,setspace}

\usepackage{amstext}
\usepackage{url}

\title{The angular nature of road networks}

\author[1,*]{Carlos Molinero}

 \author[2]{Roberto Murcio }
 
  \author[1]{Elsa Arcaute }

\affil[1]{\small{Centre for Advanced Spatial Analysis (CASA), UCL}}

\affil[2]{\small{Consumer Data Research Centre (CDRC), UCL}}

\affil[*]{c.molinero@ucl.ac.uk}

\begin{document}
\maketitle
\begin{abstract}
Road networks are characterised by several structural and geometrical properties. The topological structure determines partially the hierarchical arrangement
of roads, but since these are networks that are spatially constrained, geometrical
properties play a fundamental role in determining the network's behaviour,
characterising the influence of each of the street segments on the system.
In this work, we apply percolation theory to the UK's road network using
the relative angle between street segments as the occupation probability. The appearance of the spanning cluster is marked by a phase transition, indicating that the system behaves in a critical way.
Computing Shannon's entropy of the cluster sizes, different stages of the
percolation process can be discerned, and these indicate that roads integrate
to the giant cluster in a hierarchical manner.
This is used to construct a hierarchical index that serves to classify roads
in terms of their importance. 
The obtained classification is in very good correspondence with the official
designations of roads.
This methodology hence provides a framework to consistently extract the main
skeleton of an urban system and to further classify each road in terms of
its hierarchical importance within the system.
\end{abstract}

\section{Introduction}

The search for a science of urban processes has generated growing interest
from many different perspectives \cite{batty2013new,bettencourt2007growth,healey2006urban,wilson2012science,barthelemy2016structure}.
One field that has particularly attracted attention, is the study of road
networks, one of the most prototypical and studied network types \cite{strano2012elementary,barthelemy2013self}. Road networks
condense in its intricate configuration a countless number of interventions
which result from a myriad of historical and political micro-decisions. These
have paved the way to a hierarchical structure that can be revealed if the
system is studied as a percolating process \cite{Arcaute2015}.

Many physical processes occurring in nature can be explained as a percolating
phenomenon \cite{Stauffer1994}. As a consequence, percolation theory has
found a wide range of applications: e.g. for oil extraction \cite{King1999};
for the study of the electrical conductivity of materials \cite{Clerc1990},
of polymerization processes \cite{Coniglio1979}, of fire spreading \cite{Christensen1993},
of epidemiology \cite{Newman2002b}, and of other health aspects such as obesity
\cite{Gallos2012collective}; and to understand the modular and integrated
structure of brain networks \cite{Gallos2012small}.
In more technical terms, percolation theory aims to study how geometrical
microscopical properties affect the macroscopic configuration of the ensemble.
These types of processes present a phase transition at a specific occupation
probability, above which an infinite cluster is formed (over a theoretical
infinite lattice). 
Below this critical probability only finite clusters are generated.

In the literature there are several applications of percolation to spatially
constrained networks. Some efforts have focused on Erd\H{o}s-R\'enyi networks
\cite{Schmeltzer2014,Li2011}; others use percolation as a means to investigate
the robustness of the network \cite{Danziger2014}; and in some cases the
emergence of regions \cite{Fluschnik2014}. In addition, there already exists
an approach to determine the hierarchy of main and secondary connections
using percolation over the minimum spanning tree (MST) of a network \cite{Wu2006}.
This last approach performs well on the specific graphs studied in the paper
(Erd\H{o}s-R\'enyi, scale-free and grids), but it is not applicable to road
networks since the main premise of the paper (that the MST contains the main
roads) does not hold in road networks (the MST contains only parts of the
main roads, leaving some or, depending on the case, even most segments of the highways out of the final graph).
Within road networks specifically, percolation has been applied as an example
to look at traffic behaviour \cite{Li2015}.

In previous work\cite{Arcaute2015}, we considered the metric distance between intersections
of roads as the threshold along which the percolation process occurred. In
this work we propose a novel methodology, where the relative angle between
road segments corresponds to the occupation probability in the percolation
process. 
Relative angles between road segments have been part of the literature on
road network analysis for a long time. 
These have been used to classify different cities' typologies \cite{Strano2013,Chan2011,Barthelemy2013}
and also, they have been used to generate different representations of road
networks. In \cite{Porta2006,Masucci2016,viana2013simplicity}, the relative  angle between segments
is used to generate the dual informational graph which is then use to characterise the network behaviour.
Moreover, a whole discipline, \emph{Space Syntax}, has emerged from looking
at road networks through centrality measures which are based on relative
angles, and which serve to infer route choice, and urban structure \cite{Hillier1999,Turner2001,Serra2013}.

Exploring the road network as an angular percolation process allows us to describe the system as a critical phenomenon. In the next section we show that we can identify different growth regimes, from which we can extract the main skeleton of the network, and devise a classification for each of the road segments. The methodology to obtain these results is described after the Discussion.

\section{Results}

Our methodology shows that the system undergoes a continuous phase transition, and hence behaves in a critical way. This transition marks the point at which the giant cluster appears allowing us to extract the main skeleton of the underlying road network. In the supplementary information~\ref{experimentalResults}, we show that the transition takes place at a critical probability $p_c$ given at an angle of $\approx$ 45.76 degrees. Further details of the phenomenological and theoretical results are given also in the supplementary information~\ref{experimentalResults}, where we compute the critical exponents for the system, and introduce the corrections for the finite size effects.

Our approach shows that by applying Shannon's entropy to the distribution of cluster sizes we can distinguish several growth regimes in the percolation process and that the phase transition can be used to determine the point at which to extract the main skeleton of the network. Lastly we use these measures to construct a hierarchical index for each of the road segments a comprehensive methodology for the hierarchical classification of road networks which is $O(n)$.

\subsection{Growth regimes emerging from the angular percolation of road networks}
\label{relevantRegimes}
 
  \begin{figure}[t]
\center{\includegraphics[width=.7\linewidth]{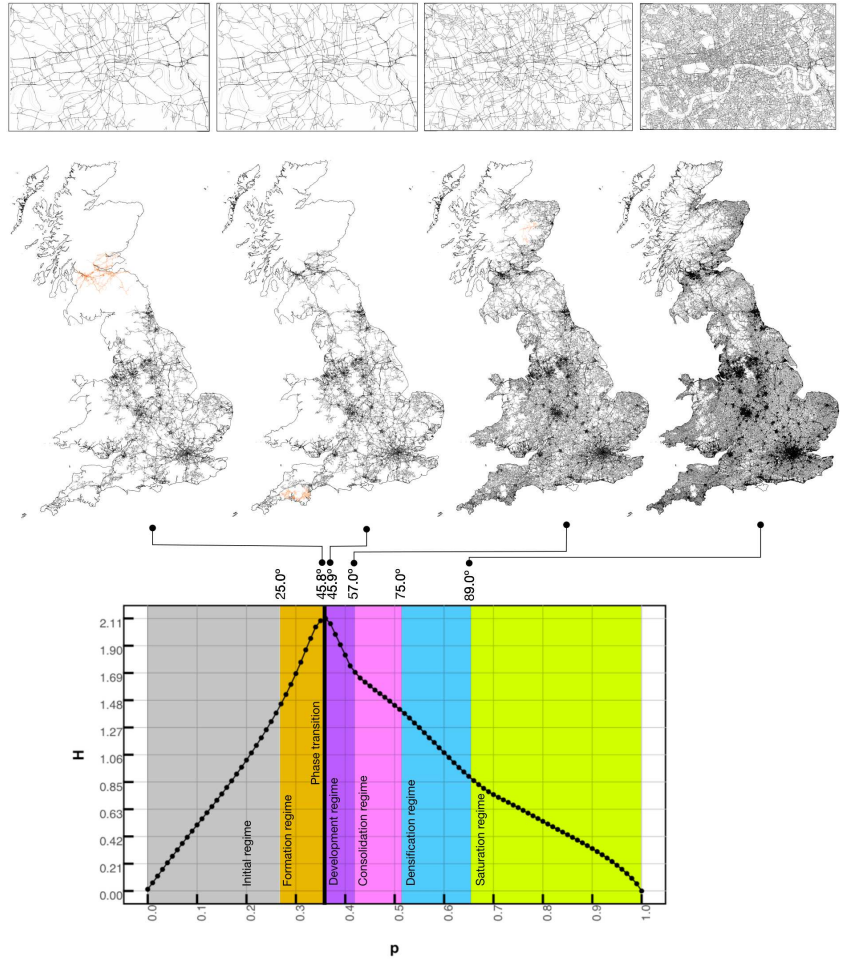}}
\caption{Shannon's entropy of the distribution of the cluster sizes. We can
clearly see 6 different regimes in the
way the system behaves. Upper panel, zoom into the center of London. Center
panel, giant cluster in black, second largest cluster in orange. Lower panel,
entropy measurement and the different regimes.
}\label{secondPhaseTransition}
\end{figure} 
 
Different growth rates can be identified as the percolation threshold is
increased. The initial regime marks an increase in the spanning capacity
of the cluster, and then shifts to densify the clusters as the probability
reaches the largest thresholds.

To determine  the regimes that characterise the formation of the giant cluster,
we analyse Shannon's entropy of the distributions of the cluster sizes at
the different thresholds
\begin{equation}
H_j=-\sum_{\forall i} p_{i,j}\cdot \log(p_{i,j}),
\end{equation}
where $H_j$ is the entropy at threshold $j$, $p_{i,j}=\frac{\sum M_{i,j}}{M}$
is the fraction of the total mass of clusters of state (mass) $i$ at threshold $j$ over the total
mass of the system. Evidently, our phase space holds all the possible states
(masses) that our system can take, including clusters with mass 0 (sites that do not belong to any cluster).

Plotting $H_j$ against the probabilities of each threshold $j$ (Figure~\ref{secondPhaseTransition}),
we observe that we can relate different speeds of variation in the entropy
levels of the distribution of the clusters, to the different slopes of the
curve. 
The change in slope determines the boundaries for the different regimes in
terms of its percolation threshold. We identify 5 different regimes:
\begin{enumerate} 
\item The initial regime in which the size of the giant cluster can be disregarded
and only very small clusters are formed.
\item The formation regime, this corresponds to the formation and growth
of the giant cluster, which will become infinite at the phase transition,
and span over the whole system (from Scotland to the southern part of England). The maximum entropy is reached at the phase transition, by the end of this regime.

\item The development regime  which starts at the phase transition at 45.763
degrees and ends approximately at 57 degrees. In this regime the giant cluster
spans the whole UK incorporating the most important roads. 
\item By the end of the consolidation regime all A and B roads with the most important minor roads (those with a dual carriageway) are incorporated to the giant cluster. This regime ends at approximately 75 degrees.

\item The next regime is the densification of the giant cluster, this marks
the beginning of the incorporation of local roads to the network. Once this
phase ends, at approx. 90 degrees, the most important local roads have also
been included. 
\item The last regime corresponds to the saturation regime, where the rest
of the local roads and alley-ways are included into the giant cluster.
\end{enumerate}

\begin{figure}[t]
\center{\includegraphics[width=1\linewidth]{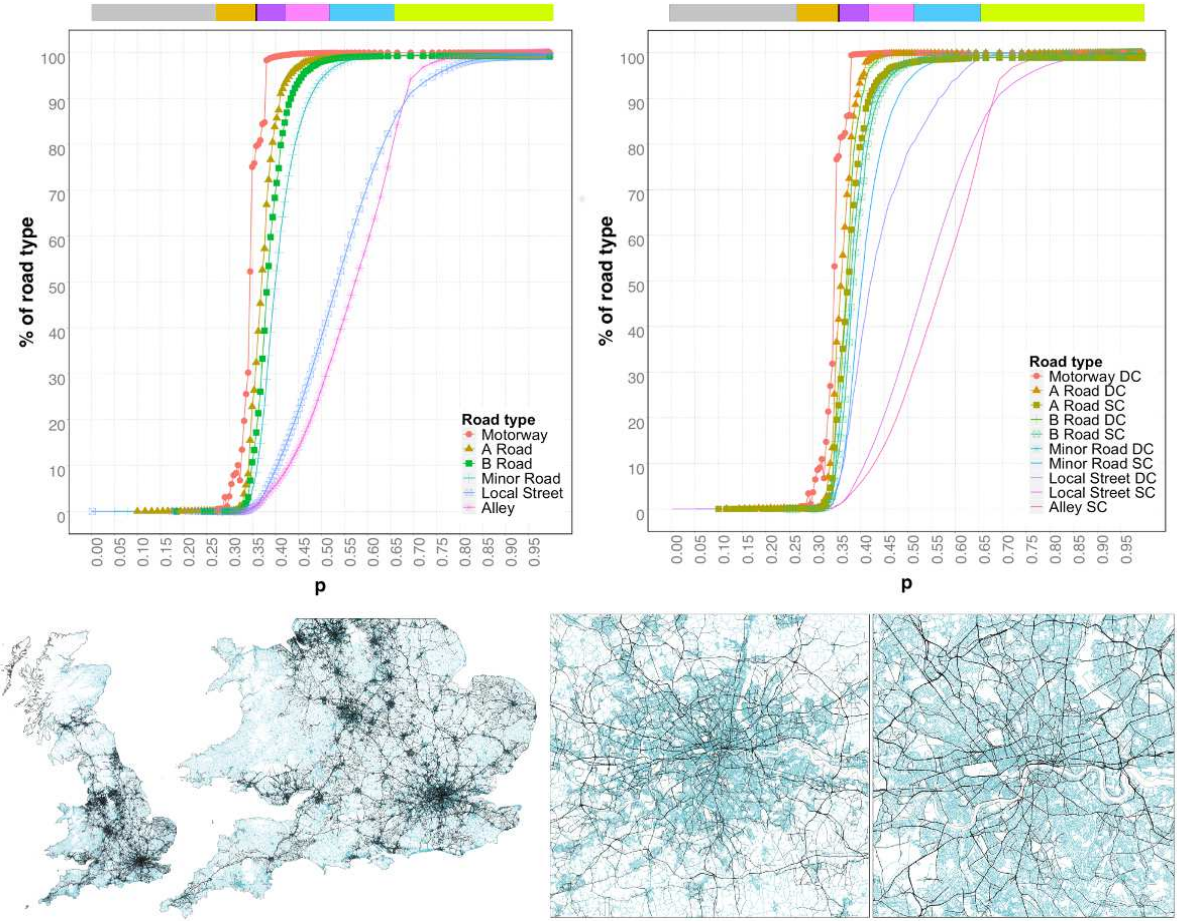}}
\caption{Upper panel, percentages of roads by road type that belong to the
giant cluster at the different percolation thresholds (in right panel, DC:
dual carriageway, and SC: single carriageway), the colors on top of the plots
represent the regimes obtained with the entropy. Lower panel, giant
cluster just after the phase transition (in black) overlaid to the full set
of roads of the road network (in light blue).
}\label{percentageRoads}
\end{figure}

The physical process taking place during the angular percolation is illustrated
in Figure~\ref{percentageRoads}.
As  the percolation threshold increases, the giant cluster sequentially incorporates
the roads by importance. At the phase transition, the main skeleton of the
road network appears. This contains the main important street segments: motorways,
A, B and minor roads. After the transition, local roads and alleys get progressively
incorporated to the giant cluster. The differentiation of importance of each
street segment is further marked by the speed of integration into the giant
cluster.

 A very interesting observation is that when the roads are disaggregated
by its secondary classification into dual/single carriageways (DC/SC), we
can see (right panel of Figure~\ref{percentageRoads}) that the different
regimes identified using the entropy analysis
actually correspond to different stages in terms of the types of roads that
are incorporated to the giant cluster. At the phase transition the percolation
incorporates to
the giant cluster all the motorways. The end of the development regime corresponds
roughly to the moment where not only the motorways but also A roads (DC)
and B roads (DC) are completely incorporated into the giant cluster. The
end of the consolidation regime corresponds to the point where simultaneously
the A roads (SC), B roads (SC), and the most important minor roads (DC) are incorporated into the cluster. By the end of the densification regime also the rest of the minor roads (SC) and the most important local roads (DC) belong to the giant cluster. The end of the saturation
regime corresponds to the inclusion of the rest of the roads (local roads
(SC) and alleys) into the giant cluster.

\subsection{Identifying the skeleton of the network}
\label{extractSkeleton}
 
The roads that are in the giant cluster just after the phase transition (which
happens at 45.763 degrees, see the table of the appendix \ref{criticalExponents}) are portrayed in black in the lower panel of Figure~\ref{percentageRoads}, while the full set of roads are in light blue.  It is easy to see that the giant cluster after the phase transition contains the main roads of the system.

Let us quantify this by looking at the percentages of roads included.
Just after the phase transition, despite the fact that the giant cluster has a
mass of only 17\% of the full road network, it already contains 98.3\% of
the Motorways, 66.9\% of the A roads, 47.7\% of the B roads, 28.8\% of the
minor roads while only containing 0.50\% of the local roads and 0.35\% of
the alleys. 
This shows that after the phase transition, the giant cluster corresponds
to the skeleton of the road network, containing the major roads.

Furthermore, if we disaggregate the roads by their secondary classification
(single or dual carriageway, SC/DC) we observe that after the phase transition
the giant cluster contains 99.5\% of the Motorways (DC), 81.5\% of A roads(DC),
71\% of B roads (DC) and 52.4\% of the minor roads (DC). 
Our approach creates a hierarchical division of the roads, from which we
can derive in a natural way the skeleton of the network without having to
use the road classification. 
The skeleton is obtained by extracting the giant cluster at the consolidation
phase, as close as possible to the critical probability in order to include
the minimum number of roads.

\subsection{Hierarchical index for road segments}
\label{hierarchicalIndex}

\begin{figure}[t]
\center{\includegraphics[width=1\linewidth]{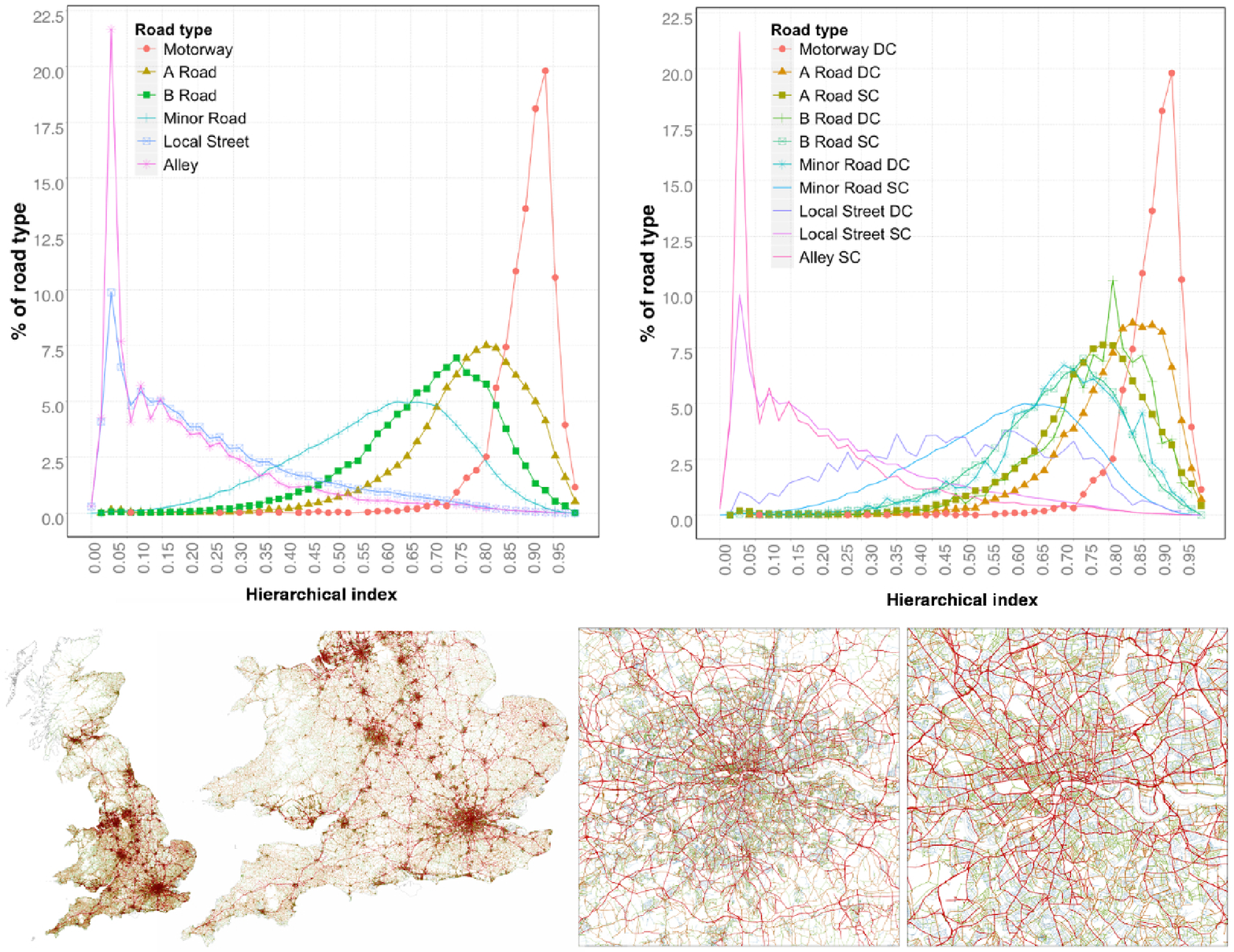}}
\caption{UK's road network at different scales where the thickness of the
lines and the color scale correspond to the values of the hierarchical index
as calculated using equation~\ref{eq::hierarchicalIndex}.
}\label{fig::hierarchicalIndex}
\end{figure} 

The hierarchical classification of road networks is of fundamental importance
to establish the routes with the highest probability to render a fast connection
between nodes.
Moreover, this is used for navigational purposes,  to aid drivers in the
identification of the most probable routes between
destinations \cite{roadClassification}. 
For this reason, several algorithms for the detection of shortest routes
in road networks have been developed, and are highly dependent on the classification
and the hierarchical organisation of the road networks \cite{Geisberger2012}.
On top of this when comparing road networks from
different countries we are bound to find  different classification systems
rendering the quest to establish an equivalence between them extremely difficult.
There is, therefore, a large interest in the generation of a methodology
that automates the hierarchical classification of road networks.

Centrality measures have played a major role in the description of systems
of road networks allowing to uncover a large number of its properties. 
As such, one of the most accepted methodologies to visualize the hierarchical
organisation of the road network is the use of betweenness centrality \cite{Freeman1977}.
This gives a value for each road proportional
to its flow through the system.
A large body of research has been devoted to the analysis of flows using
centrality measures \cite{Borgatti2005,HOLME2003}
and, in the case of road networks, a large part of that body of research has been devoted to the use of
angular distances \cite{Dalton2003,Turner2001}, which seems to improve the
detection of these main flows \cite{Hillier2005}.
The large drawback of using betweenness to generate an index is that the
complexity of its computation  ($\approx O(n^3)$, where $n$ is the number
of nodes of the graph) makes infeasible its use for large systems. Another methodology that we can find to determine the hierarchy of reticular networks using a graph theoretic approach is the one found in \cite{mileyko2012hierarchical} that studies its application in biological networks.

In this section we propose an alternative methodology to generate a hierarchical
index for each road segment that is based on the percolation process. This
technique is computationally less demanding than calculating the betweenness
index, and it is linear in complexity ($\approx O(n)$) which means that it
is fully capable of analysing large systems. We construct the index according
to the following principle. This consists in assigning a degree of importance
to the street segment according to its contribution to the informational
content of the system. To do this, we weight the entropy of the system with
the mass of the cluster to which the street segment belongs. Its total contribution
needs to be considered for each percolation threshold, hence we sum over
all the thresholds. Given that the cluster sizes follow a power law distribution,
it is more appropriate to weight the entropy with the log of the mass insted
of the mass itself. More formally we can write that the normalised hierarchical
index $I_i$ of road $i$ is:

\begin{equation}
I_i=\frac{ \sum_{j=1}^t (H_j\cdot \log(M_{i,j}))}{\sum_{j=1}^t (H_j\cdot
log(max(M_{j})))}
\label{eq::hierarchicalIndex}
\end{equation}
where $j$ runs through all the percolation thresholds ($t$), $M_{i,j}$ is
the mass of the cluster that contains road $i$ at the $j$-{\it th} threshold,
$H_j$ is the entropy of the distribution of the cluster sizes at the $j$-{\it
th} threshold and $max(M_{j})$ is the maximum mass of all clusters at threshold
$j$. The denominator is just a normalisation constant.

The results of our index are shown in Figure~\ref{fig::hierarchicalIndex}.
We can see that roads have been assigned an index that is consistent with
the given road classifications (the highest values correspond to Motorways,
then A-roads, followed by B-roads, minor roads, local streets and alleys).
For further granularity, we look at the sub-classifications differentiating
between DC and SC. Observing the right upper panel, we can see that the index
classifies with the same histogram A-roads (SC) and B-roads (DC), and the
same holds true for B-roads (SC) and minor roads (DC). Meanwhile, A-roads
(DC) get closer to their classification as Motorways. 
These observations hint that though their classification is a priori different,
the fact that internally (in the comparison between the secondary classification
SC/DC) dual-carriageway roads are more important than single ones affects
their hierarchical position within the system.

\section{Discussion}
\label{conclusions}

Throughout this work we have shown that the road network viewed as a percolation
process
behaves in a critical way. The phase transition marks a differentiated growth
rate, and the threshold at which the giant component can be identified with
the skeleton of the road network. Furthermore, we observed a hierarchical
organisation on the importance of the road segments with respect to their
contribution to the entropy of the system at each threshold of the percolation.
This was used to construct a hierarchical index for each road segment that
serves to classify the different roads in the network.

In conclusion, we have shown that the road network encounters a natural description
when seen as a percolation process, where the threshold corresponds to the
angle between the street segments. These are very promising results that
open up new possibilities for further studying
the road network properties under this paradigm. Furthermore, the hierarchical
index can serve as the basis for an algorithm in the spirit of contraction
hierarchies that could improve the speed of shortest path analyses. These
are properties that cannot be observed in the system under its metric description
and that, therefore, highlights the angular nature of the road network.

A description of the phenomenon in terms of the critical exponents can be found in the appendix (\ref{experimentalResults}) where we also propose the corrections for the exponents taking into account the finite size effects. In future work, we intend to apply this process to other systems, such as road networks from
different continents and natural systems, in order to establish a basis for
comparison through the critical exponents of the infinite lattice. This will
give us insight into whether these systems share generic properties, or whether
a classification can be achieved; in addition to investigate whether growth
processes can be inferred from these behaviours.

\section{Methods}

\label{methodology}

The system under consideration is the UK road network.
The original data is obtained from the OS MasterMap database  (ITN layer)
\cite{MasterMap2010}. 
The layer is processed and simplified as follows: lanes are collapsed into
a single link without considering directionality, and the nodes of degree
2, which correspond to intermediate points  are removed. Only nodes corresponding
to intersecting roads are kept.
Note that by removing these nodes more emphasis is put on the structural
properties of the network, since the angles within a road are not considered.

The properties of road networks can be divided into two subclasses: structural
and geometrical. 
The former are studied by representing the network as a graph, where the
nodes are the road intersections and the links are the street segments. 
This allows us to calculate many network features: e.g. the degree distribution,
centrality measures, the spectrum of the graph, or to extract communities,
among many.
Geometrical properties, in turn, are related to distances over the network,
to lengths of the street segments, the width of roads, the slopes formed
by the topography,  the relative angle between two street segments, and so
forth. 
One of the most common ways to incorporate these geometric factors into the
analysis is to include them as weights in the graph of the road network.

Note that these weights refer to a relationship between the nodes in the
graph, as is the case of the length of roads, but in our case, the relative
angle is a relationship between the links of the graph (the road segments).
In order to account for this second type of relationship, we need to generate
the link-node dual (or line-graph \cite{Criado2011}) of the road network's
graph.

\begin{figure}[t]
\center{\includegraphics[width=.85\linewidth]{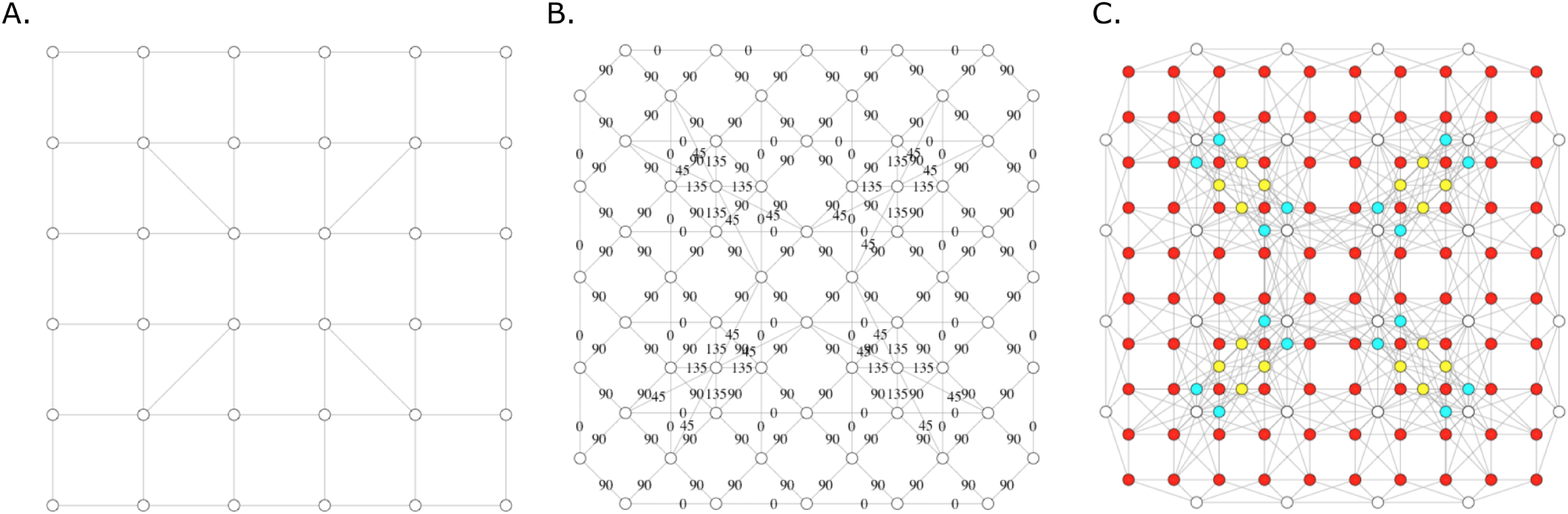}}
\caption{(a) example of a primal graph of a street network ($G$), (b) line-graph
generated from that primal graph ($G'=L(G)$), (c) line-graph of the line-graph
where colors represent different probabilities for each site ($G''=L(L(G))$).}
\label{p}
\end{figure} 

Let us denote by $G$ the graph of a road network  (Figure~\ref{p}.a), its
line-graph, denoted by $G'=L(G)$, is constructed as follows (Figure~\ref{p}.b).
Each link (street segment) in the original graph, is replaced by a node in
the line-graph, and a link is created in the line-graph if two links of $G$
share a node.
The line-graph generated from the network holds 4 million nodes and 7 million
links after processed and simplified.
The relative angle between road segments can now be mapped to the weights
of the line-graph.
This procedure is very similar to the one presented in \cite{Turner2001},
although here we do not normalize the values of the relative angles.

The bond percolation is then executed on the line-graph ($G'$) of the street
network as follows.
Given a certain angular threshold, the occupation probability for a link
is the  probability that a link of the line-graph has a weight equal or below
the angular threshold (Figure~\ref{p}.c).
We can then associate a probability to the angular threshold by computing
the percentage of links that are below or equal to this threshold. For example,
the probability associated to 45 degrees is $p=0.376$ since 37.6\% of the
links are below or equal to 45 degrees (see Figure~\ref{firstPhaseTransition}.b).

Taking into account that what we are calculating is a bond percolation on
the line-graph, we can consider that conceptually we are performing a site
percolation on the line-graph of the line-graph (Figure~\ref{p}.c) where
every node will correspond to a site of the lattice used for the percolation,
and the probabilities of each site to be occupied are proportional to the
weights of the line-graph. This $L(L(G))$ will serve us as the equivalent
of the lattice in a typical percolation process and we will use it to calculate
the fractal dimension of our system. Throughout the text we will refer to
the number of links in the line graph as the mass of the system (or say of
a particular cluster), which is equivalent to the number of nodes of the
$L(L(G))$.

The interested reader can find all the specificities of the algorithm to
calculate the percolation process in the section of the appendix~\ref{algorithm}.

\section{Appendix}

\subsection{Percolation theory and phenomenology}
\label{experimentalResults}

\begin{figure}[t]
\center{\includegraphics[width=1\linewidth]{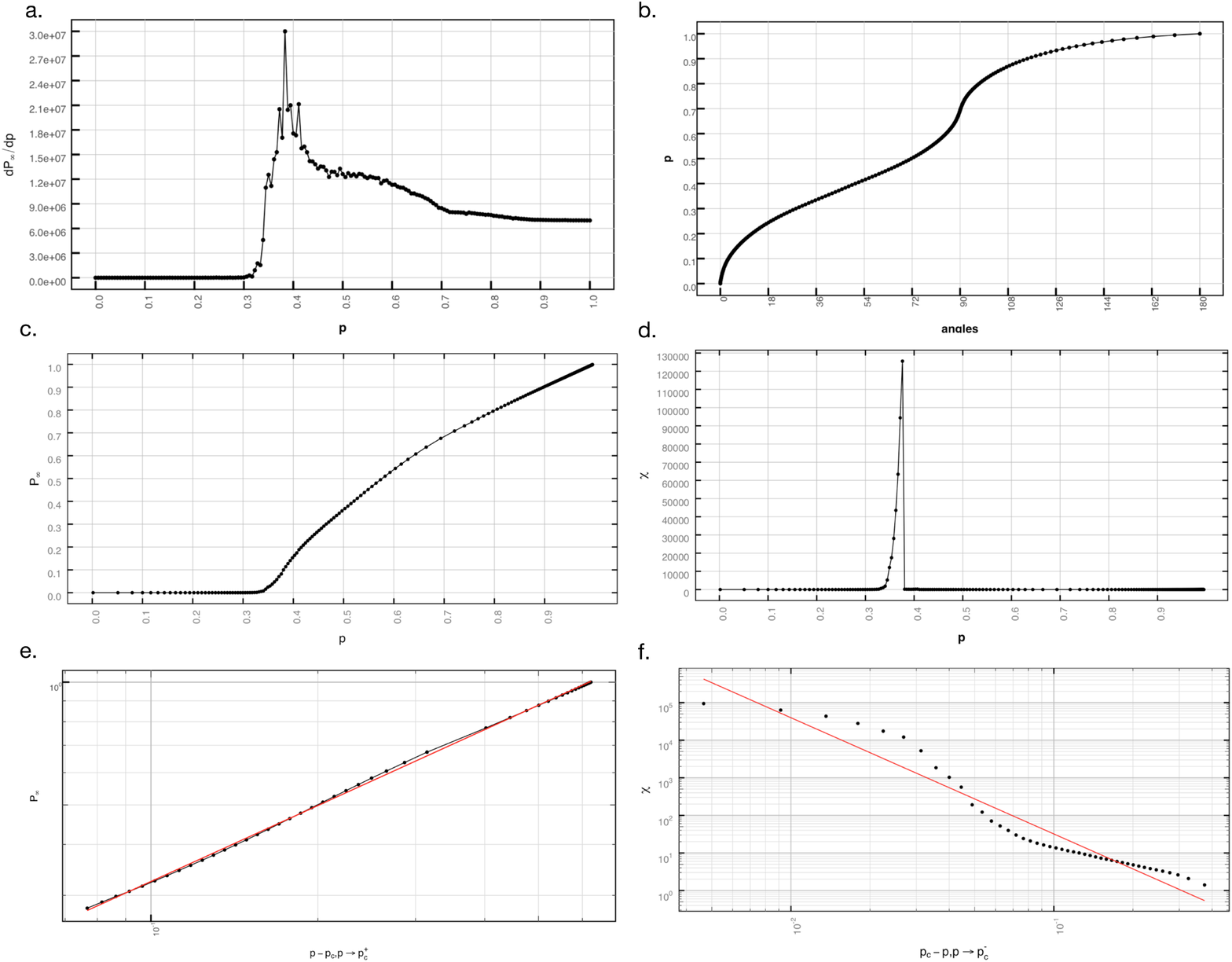}}
\caption {Critical behaviour of the UK road network: a) derivative of the
order parameter; b) relationship between each angle and its cumulative probability
of occurence; c) order parameter $P_\infty$; d) average cluster size $\chi$;
e) critical
exponent $\beta$; and f) critical exponent $\gamma$.}\label{firstPhaseTransition}
\end{figure} 

Percolation processes are critical phenomena that present continuous phase
transitions that can be characterised by critical exponents \cite{Christensen2005}.
In this section we show how these exponents can be described and computed
for our system. 
We correct for any errors arising from finite size effects in our calculations
through the introduction of a series of subsystems of different sizes.

\subsubsection{Critical exponents}

Critical phenomena have been largely studied in physical systems, in particular
in systems where the temperature $T$ defines the various phases, and the
critical point at which the transitions take place \cite{meakin1998fractals}.
For these systems in general, a full ordered system can be found at $T=0$,
and as this is increased, it reaches a critical point $T_c$ at which the
system undergoes a continuous phase transition, and the correlation length
diverges $\xi \to \infty$. As the temperature is increased, at the asymptotic
limit $T \to \infty$ the system is fully disordered.
In percolating systems, the occupation probability takes the role of the
temperature, and at a critical probability $p=p_c$, the system undergoes
a second order phase transition and a giant cluster appears \cite{Stauffer_Aharony_percolation1994}
spanning the whole space. At this point the percolation clusters become self-similar,
and the system can be described as a fractal \cite{Bunde_Havlin_Fractals}.
This transition can be fully characterised in terms of critical exponents
which only depend on the dimension of the space and type of percolation \cite{Stauffer_Aharony_percolation1994}.
Although several exponents can be defined, only two of them are independent,
and the rest can be obtained through a series of scaling laws. 

Cities are known to have a fractal structure as demonstrated a
few decades ago by \cite{frankhauser1998fractal,Batty_LongleyFractasl1994},
and more recently they have been considered as multifractals \cite{Murcio_Multifractal2015,chen2013multifractal}.
We will hence consider that our system (the road network) is a fractal and
define the  fractal dimension $d$ of our system as the capacity dimension
of the $L(L(G))$
of the road network graph. This fractal dimension will be measured by using
the box-counting methodology on the nodes of the graph $L(L(G))$.
This allows us to establish the relationship $M=l^d$,
where $M$ corresponds to the mass of the system, in this case the number
of nodes of L(L(G)) (equal to the number of links of the L(G)), and $l$ to the theoretical lattice length. This
equivalence allows us to calculate a theoretical lattice size (which will
be needed to remove finite size effects later on)
\begin{equation}
l=M^{\frac{1}{d}}\label{l}
\end{equation}

At the percolation threshold $p_c$, the giant component spans the whole system
and becomes self-similar \cite{Christensen2005,Stauffer_Aharony_percolation1994,Bunde_Havlin_Fractals}.
The fractal dimension $D$ of the largest component at the percolation threshold
can be obtained in a similar way
\begin{equation}
M_\infty (p_c,l)= l^D\label{M_D}
\end{equation}
where $M_\infty(p_c,l)$ is the mass of the spanning cluster at the phase
transition, and $l$ is the size of the lattice. Moreover, since $P_\infty=\frac{M_\infty}{M}$
we can say $P_\infty(p_c,l)=\frac{l^D}{l^d}$ or $P_\infty(p_c,l)=l^{D-d}$.
Equation~\ref{M_D} can be used to calculate $D$ given that we can directly
measure $M_{\infty}$ and we can calculate $l$ (equation~\ref{l}) then $D=\frac{log(M_\infty)}{log(l)}$.

In a typical percolation process, the probability of a site belonging to
the giant cluster $P_\infty=\frac{M_\infty}{M}$ (Figure~\ref{firstPhaseTransition}.c)
takes the role of the order parameter. It is practically 0 below the phase
transition and increases rapidly after reaching a fully ordered system at
$p=1$. 

This quantity will allow us to calculate the critical probability $p_c$ at
which the system undergoes a phase transition. In order to find the location
of the phase transition we  detect the maximum of the derivative of our order
parameter as shown in Figure~\ref{firstPhaseTransition}.a (another methodology
to calculate $p_c$ would be to look for the threshold that maximises the
mass of the second largest cluster). 
For the UK road network we obtain $p_c$ at an angle of $45.7631\pm0.0001$
degrees, which corresponds to the probability $p_c=0.3783$.

The behaviour of  $P_\infty$ at the critical point is characterised by the
critical exponent $\beta$ (Figure~\ref{firstPhaseTransition}.e) as follows
\begin{equation}
 P_\infty\propto \vert p-p_c\vert ^\beta, p\rightarrow p_c^+
 \label{eq_Pinf}
 \end{equation}

The clusters that appear at the different probabilities $p<p_c$ are characterised
by their linear dimension $\xi$ \cite{stanley1999scaling}. This is one of
the most important variables in critical phenomena, containing the information
of the range of correlations given by the interactions. At the phase transition,
the correlation length diverges, $\xi \to \infty$. Its behaviour close to
the critical point is also a power law leading to the critical exponent $\nu$
\begin{equation}
\xi\propto\vert p-p_c\vert ^{-\nu}, p\rightarrow p_c 
\label{eq_correlationLength}
 \end{equation}

Let us look at other important quantities in percolating processes that give
rise to these exponents.
The clusters that appear at the different probability thresholds can be characterised
according to their average size $\chi$. 
This is defined as follows
\begin{equation}
\chi =\frac{ \sum s^2 n_s}{\sum s n_s}
\end{equation}
where $s$ is the cluster size and $n_s$ the normalised number of clusters
of size $s $ per site (the number of clusters of size $s$ normalised by the
total number of sites in the lattice). 
The average cluster size increases as $p$ increases, until the critical probability
$p_c$ is reached. Once this happens, the giant clusters spans the system,
and the average cluster size drops suddenly, since above the critical probability
the giant cluster is  infinite and therefore is not taken into account, see
Figure~\ref{firstPhaseTransition}.d.
The critical exponent $\gamma$ associated with $\chi$ hence measures the
increase and decay of the size of the finite clusters around the critical
probability $p_c$, and this is obtained from the following equation
\begin{equation}
 \chi\propto \vert p_c-p\vert ^{-\gamma}, p\rightarrow p_c \label{chi_gamma}
\end{equation}
see Figure~\ref{firstPhaseTransition}.f.
The typical size of the largest cluster $s_{\xi}$ is referred as the characteristic
cluster size, and it can be obtained in a similar way
\begin{equation}
s_\xi\propto \vert p_c-p\vert ^{-\frac{1}{\sigma}}, p\rightarrow p_c
\end{equation}
where the exponent $\sigma$ determines the speed of the variation of the
characteristic cluster size.
Lastly, the Fisher exponent $\tau$ characterizes the variation of the normalized
number of clusters around $p_c$.

\begin{equation}
n_s\propto s^{-\tau}\cdot \mathcal{G}(s/s_\xi)
\end{equation}
where $s$ is the size of the cluster and $s_\xi$ is the characteristic cluster
size and $\mathcal{G}(s/s_\xi)$ is the scaling function for the cluster number
density and is defined as $\mathcal{G}(s/s_\xi)=(s/s_\xi)^2\cdot exp(-s/s_\xi)$.

These exponents are related to each other according to some scaling laws, and from all this collection, only two are independent. We can therefore use these scaling laws to analytically calculate the rest of the exponents once we have measured two of them. The scaling laws derived from taking $D$ and $\beta$ as the independent exponents
are :
\begin{eqnarray}
\nu&=&\frac{\beta} {d-D}\\ \label{scaling1}
\gamma&=&\nu (2D-d) \label{scaling2}
\end{eqnarray}
which lead to the well-known variation of Rushbrooke inequality \cite{stanley1999scaling}
\begin{equation}
2\beta+\gamma=\nu d \label{2beta_gamma}
\end{equation}
In terms of $\sigma$ and the Fisher exponent ($\tau$), we have
\begin{eqnarray}
\sigma&=&\frac{1}{\nu D}\label{sigma} \\
\tau&=&\beta \sigma +2 \label{tau}
\end{eqnarray}
The initial results are given in Table~\ref{criticalExponents} under the
column ``Finite system''. These results do not take into account corrections
that need to be introduced given the finite size of the system.
\begin{table}[h]
\centering
\label{my-label}
\begin{tabular}{|l|l|l|}
\hline
Exponent & Finite system&Infinite system \\
\hline
$\beta$  & 0.6186(*)&0.6187(*)\\
$\gamma$ & 2.8100 &2.9053(*)\\
$\nu$    & 2.2182&2.2706(*) \\
$\tau$   &2.1804 &2.1763\\
$\sigma$ & 0.2917& 0.2849\\
$D$     & 1.5456(*)&1.5456 \\
\hline
$d$     & 1.8245(*)&1.8245\\ 
\hline
$p_c$     & 0.3783(*) [45.763$^o$]&0.3766(*) [45.39$^o$]\\
\hline
\end{tabular}
\caption{Critical exponents obtained for the UK road network: before (Finite
system), and after introducing corrections (Infinite system) due to the finite
size of the system. 
The values marked with (*) are the ones that have been calculated from the
experimental results, the rest of the exponents were calculated by using
the scaling relationships between the exponents.\label{criticalExponents}}
\end{table}
In the following section we introduce these.

\subsubsection{Removing the finite size effect}

\begin{figure}[t]
\center{\includegraphics[width=1\linewidth]{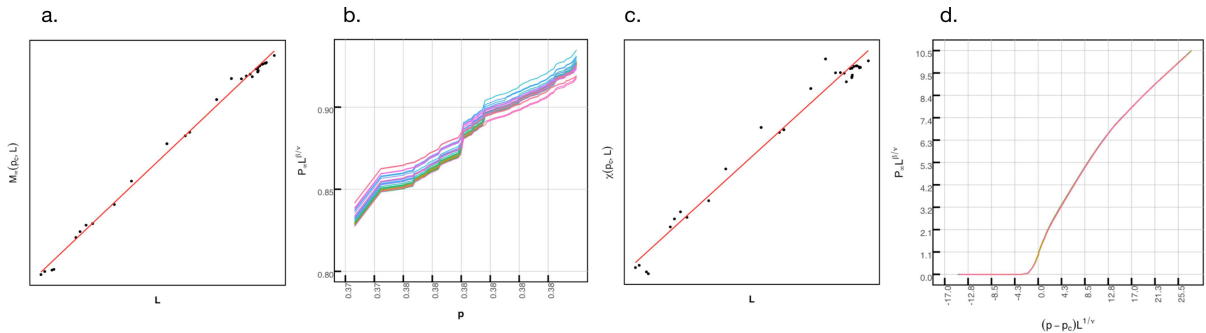}}
\caption{ Calculations correcting for the finite size effect (infinite
theoretical system), from left to right, (a) calculation of $D$, (b) correction
of $\frac{\beta}{\nu}$ and simultaneously determination of  $p_c$,  (c) calculation
of $\frac{\gamma}{\nu}$, (d) data collapse to calculate $\nu$.}\label{firstPhaseTransitionFSE}
\end{figure} 

Correcting for finite size effects is not a trivial matter. One methodology
that can be employed \cite{cardy1996scaling} is to consider different sizes
of the system and perform the analysis considering that the results are valid
for the asymptotic limit, which in this case corresponds to having the size
of the system much larger than the correlation length. Effectively we are
rescaling the results, so we obtain a data collapse.

We generate 32 subsets of different size of road networks from the original
one.
Each of those systems will have a different mass and lattice size, but will
maintain the same fractal dimension.
Following equations (\ref{eq_Pinf}) and (\ref{eq_correlationLength}),  we
can write $P_\infty$ as a function of the correlation length at the asymptotic
limit
\begin{equation}
P_\infty\propto\xi^{-\frac{\beta}{\nu}}, p\rightarrow p_c^+
\end{equation}
At $p_c$ the correlation length $\xi$ becomes larger than any of the lattice
sizes of our finite systems, and the equation no longer holds. We solve this
by capping the size of the largest cluster at $p_c$  by the lattice size
as follows: $P_\infty(p_c,l)\propto l^{-\frac{\beta}{\nu}}$, which is equivalent
to: $P_\infty(p_c,l)\propto l^{D-d}$. Therefore, we can obtain an initial
estimate for $\frac{\beta}{\nu}$ by taking a measure of the slope of a log-log
plot of $M_\infty$ against $l$ for all the subsystems (Figure~\ref{firstPhaseTransitionFSE}.a)
which gives us an estimate of $D$ (equation~\ref{M_D}), that can then be
inserted into $-\frac{\beta}{\nu}=D-d$.   

We should remark that the critical probability in an infinite lattice will
probably be different from the previously calculated one in our finite system.

In order to find $p_c$ for the infinite case, we use the methodology proposed
in \cite{Tsakiris2010}.
This consists in plotting $P_\infty(p_c,l)\cdot l^{\frac{\beta}{\nu}}$ vs.
$p$, and adjusting the value of $\frac{\beta}{\nu}$ until all the curves
cross in one single point (Figure~\ref{firstPhaseTransition}.b). That point
will be our $p_c$ and the final value of $\frac{\beta}{\nu}$ will be given
by this procedure.

In a similar manner, using equations (\ref{eq_correlationLength}) and (\ref{chi_gamma}),
and capping the size of the largest cluster at $p_c$, we can calculate $\chi(p_c,l)$
at different lattice sizes at the critical probability so that 
\begin{equation}
\chi(p_c,l)\propto l^\frac{\gamma}{\nu}
\end{equation}
allowing us to determine $\frac{\gamma}{\nu}$, see Figure~\ref{firstPhaseTransitionFSE}.c).

Let us now determine $\nu$ correcting for the finite size effect.
For all the different lattices, we plot $P_\infty\cdot l^\frac{\beta}{\nu}$
vs. $(p-p_c)\cdot l^{\frac{1}{\nu}}$ (Figure~\ref{firstPhaseTransitionFSE}.d)
and adjust the value of $\frac{1}{\nu}$ until all the data collapses. This
gives us the value of $\nu$ and with the calculated values of $\frac{\beta}{\nu}$
 and $\frac{\gamma}{\nu}$ we can obtain $\beta$ and $\gamma$. The fractal
dimension $D$ can be computed using equation \ref{scaling1}. The scaling
laws (\ref{sigma}), (\ref{tau}) can be used to find $\sigma$ and $\tau$.
The results for the ``Infinite system'' can be found in Table~\ref{criticalExponents}.

\subsection{Algorithm for the percolation on the road network}
\label{algorithm}

Given a graph of the road network $G=(V,E)$, where $V$ represent intersections
and $E$ represent streets, we generate its line-graph and assign the angular
distance between streets as the weights of the links of the line-graph $w$
($G'=(V',E',w)$
For a certain angular threshold ($t$) (e.g. $t=40$ degrees) we produce a
network percolation as follows:

\begin{enumerate}
\item Delete all links that are above the threshold.
\item Find the connected components of the resulting graph. Each of those
components will be a cluster of the percolation.
\end{enumerate}

This simple algorithm is $O(n)$, given the nature of the system. Road networks'
graphs have very specific characteristics, they are planar,
sparse and due to spatial restrictions (i.e. the width of the streets) their
maximum degree is very limited (to around $k_{max}=10$). This gives the following
results:

\begin{itemize}
\item The number of links in a planar graph is linear with respect to the
number of nodes $m=c\cdot n$ where $m$ is the number of links, $n$ the number
of nodes and $c$ a constant. 
\item The complexity of generating a line graph is given first by going through all its links to generate a node and then to generate the links by going through
all the links and then through all the links that depart from its end nodes.
 To simplify, we can say that the complexity will have an upper bound of $\Omega(m+2mk_{max})$, where here $k_{max}$ refers to the maximum degree of a node. This will be linear in road networks since in these graphs the maximum degree can be considered constant $k_{max}=c$, so its complexity is more or less $O(m\cdot c)$ and because the graph is sparse the complexity is more or less linear in the number
of nodes $O(n)$.
\item The algorithm that generates the clusters following the percolation
procedure is linear in time. The first step is linear in the number of edges
and the second step can be implemented with a BFS (breadth-first search)
which runs also in linear time ($O(n+m)$). The execution time will be of
$O(m+n+m)$
which can be reduced to $O(n)$ for sparse graphs such as road networks. So
the total execution time of the algorithm (including generating the line
graph) in road networks is $O(n)$. 
\end{itemize}
The algorithm will be executed $t$ times , being $t$ the number of thresholds, which is independent of the input ($n$). Therefore, the complexity  for executing the algorithm remains linear ($O(n)$).

\bibliographystyle{naturemag-doi}

\section*{Acknowledgements}

The authors wish to thank Remi Louf for his comments and useful insights
on the final version of the text.
The authors acknowledge the support of the ERC grant no. 249393-ERC-2009-AdG.
CM and EA also acknowledge the support of the EPSRC grant: EP/M023583/1 and RM also acknowledges the support of UK ESRC Consumer Data Research Centre (CDRC) [ES/L011840/1].

\section*{Author contributions statement}

C.M. conceived the experiment(s), conducted the experiment(s), analysed the results. All authors reviewed and contributed to the writing of the manuscript and to the development of the mathematical framework.

\end{document}